\documentclass[english,utf8x]{article-hermes_french}
\usepackage[labelfont=bf,textfont=it,labelsep=period,justification=raggedright,singlelinecheck=false]{caption}

\usepackage{url}
\frenchsetup{AutoSpaceFootnotes=false}

\journal{TAL. Volume 65 -- n°2/2024}{7}{13}

\title[Short title]{Preface to the Special Issue of the TAL Journal on Scholarly Document Processing}

\author{Florian Boudin\fup{*} \andauthor Akiko Aizawa\fup{**}  } 

\address{%
\fup{*} JFLI, CNRS, Nantes University, France \\
\fup{**}  National Institute of Informatics, Japan
}

\abstract{
The rapid growth of scholarly literature makes it increasingly difficult for researchers to keep up with new knowledge.
Automated tools are now more essential than ever to help navigate and interpret this vast body of information.
Scientific papers pose unique difficulties, with their complex language, specialized terminology, and diverse formats, requiring advanced methods to extract reliable and actionable insights.
Large language models (LLMs) offer new opportunities, enabling tasks such as literature reviews, writing assistance, and interactive exploration of research.
This special issue of the TAL journal highlights research addressing these challenges and, more broadly, research on natural language processing and information retrieval for scholarly and scientific documents.
}


\motscles{
Traitement automatique de documents scientifiques,
Traitement Automatique des Langue pour la science,
Grands modèles de langues (LLMs).
}

\resume{
La croissance rapide de la littérature scientifique rend de plus en plus difficile pour les chercheurs de suivre l’évolution des connaissances.
Le recours à des outils automatisés est aujourd’hui indispensable pour naviguer et interpréter cette immense masse d’informations.
Les articles scientifiques posent des difficultés uniques en raison de leur langage complexe, de leur terminologie spécialisée et de leurs formats variés, ce qui nécessite des méthodes avancées pour extraire des informations fiables et exploitables.
Les grands modèles de langage (LLMs) ouvrent de nouvelles perspectives, permettant des tâches telles que les revues de littérature, l’assistance à la rédaction et l’exploration interactive des travaux scientifiques.
Ce numéro spécial de la revue TAL met en lumière des recherches qui s'attaquent à ces défis et, plus largement, des recherches sur le traitement automatique des langues et la recherche d'information appliqués aux documents scientifiques et académiques.
}

\begin{document}

\maketitlepage

\section{Introduction}

The volume of scholarly literature is expanding rapidly.
A compelling example is the ACL Anthology\footnote{\url{https://aclanthology.org/}}, a repository for scientific contributions within the fields of computational linguistics and Natural Language Processing (NLP), which recently surpassed 100,000 papers, doubling its size in just four years~\cite{bollmann-etal-2023-two}.
As the rate of publication continues to accelerate, researchers and institutions face increasing challenges in keeping up with the flood of new knowledge.
This highlights the critical need for automated methods to help navigate, understand and distill the growing body of scientific information.

To address this pressing challenge, researchers across various fields---including computational linguistics, NLP, text mining, information retrieval, digital libraries and scientometrics---have dedicated significant efforts into developing methods and resources designed to process scientific documents.
This led to a surge in publications on the matter, alongside the successful hosting of numerous international events, such as the workshops \emph{Scholarly Document Processing (SDP)}~\cite{sdp-2024-scholarly}, \emph{SCIentific DOCument Analysis (SCIDOCA)}~\cite{scicoda-2024}, \emph{Natural Language Processing for Scientific Text (SciNLP)}~\cite{scinlp-2021} and \emph{Bibliometric-enhanced Information Retrieval (BIR)}~\cite{bir-2024}.

At the national level in France, scholarly document processing is also gaining momentum.
This interest is exemplified by the success of the workshop \emph{Analyse et Recherche de Textes Scientifiques (ARTS)\footnote{\url{https://arts2023.sciencesconf.org/}}}~\cite{jep-taln-recital-2023-actes-de-coria}, held at the TALN-CORIA 2023 conference.
The event, which saw the presentation of 12 papers and attracted over 40 participants, highlighted the relevance of the topic and prompted the call for this special issue of the TAL journal.

Scientific papers present unique challenges for document processing methods due to their inherent complexity.
They are characterized by intricate technical language, discipline-specific terminology, distinct structural conventions, and frequent use of mathematical expressions, all of which pose significant challenges for current methods~\cite{10.1007/s00799-023-00352-7}.
Additionally, the multi-modal nature of scientific papers, with their tables, figures and diagrams, further complicates their processing~\cite{shen-etal-2022-vila}.
Beyond these document-level challenges, effective methods should also account for features present at the collection level, such as citation networks, and leverage rich metadata, including authors, keywords, and publication venues, each introducing its own set of difficulties.

Developing methods to extract reliable, valuable and verifiable information from scientific papers is crucial for many downstream applications, including retrieval~\cite{boudin-etal-2020-keyphrase,NEURIPS2023_78f9c04b}, recommendation~\cite{kreutz2022scientific,huang2024scientific}, summarization~\cite{10.1145/3543507.3583505}, question-answering~\cite{saikh2022scienceqa,auer2023sciqa} and document understanding~\cite{wright-augenstein-2021-citeworth,veyseh2021acronym}.
With the rise of large language models (LLMs) and their enhanced ability to analyze and synthetize insights across multiple scientific papers, new applications are continuously emerging.
Promising developments include accelerating scientific discovery~\cite{zhang-etal-2024-comprehensive-survey}, generating novel research directions~\cite{wang-etal-2024-scimon}, reviewing of the literature~\cite{newman-etal-2024-arxivdigestables}, assisting scientific writing~\cite{jourdan-etal-2024-casimir} and enabling interactive exploration of papers~\cite{zheng-etal-2024-openresearcher}.

LLMs are also being developed for specialized scientific domains, such as healthcare and medicine~\cite{labrak-etal-2024-biomistral} or material sciences~\cite{zhang-etal-2024-honeycomb}.
These domain-specific models assist experts and researchers with complex tasks, including drug discovery~\cite{savage2023drug}, diagnosis generation~\cite{info:doi/10.2196/51391}, and science education~\cite{cooper2023examining}.

Efforts are also underway to reduce the growing computational and environmental costs associated with training and deploying LLMs~\cite{hershcovich-etal-2022-towards,sustainlp-2023-simple}.
At the same time, ethical concerns are being addressed, with research focused on the responsible use of LLMs in science, including issues of bias, fairness, and transparency in AI-driven research~\cite{nlp4science-2024-1}.

This special issue of the TAL journal focuses on research addressing these challenges, with an emphasis on \emph{NLP and information retrieval for scholarly and scientific documents}.

\section{Call, Reviewing and Selection of Papers}

The call for submissions to this special issue of the TAL journal on scholarly document processing was announced in December 2023, and the submission platform\footnote{\url{https://tal-65-2.sciencesconf.org/}} closed in March 2024.
The scope of relevant topics extended beyond NLP and information retrieval tasks, tools, and resources designed for scientific documents, encompassing areas such as bibliometrics, scientometrics, citation analysis and recommendation, claim verification, plagiarism detection and scientific writing assistance.

A total of five articles were submitted (two in French and three in English) by authors from Iran, India and France.
Each article was reviewed by three experts, two members of the scientific committee and one member of the Editorial Board.
In May 2024, the Editorial Board and the guest editors met to discuss the first round of reviews and notified the authors of the outcomes.
One paper was selected for a second round of reviews and was ultimately accepted, resulting in a selection rate of 20\%.

\section{Accepted paper}

This issue of the TAL journal features one paper: \emph{Évaluation de la qualité de rapport des essais cliniques avec des larges modèles de langue} (Evaluating clinical trials research article quality with large language models) by Mathieu Laï-king and Patrick Paroubek.
The paper focuses on the biomedical domain, specifically investigating the use of LLMs to evaluate the quality of Randomized Controlled Trials (RCTs), a type of clinical research article.
The authors frame the evaluation task as a question-answering problem, prompting LLMs to assess articles according to the Consolidated Standards of Reporting Trials (CONSORT) framework.
Through extensive experiments, the study demonstrates the high effectiveness of LLMs, achieving an accuracy of up to 85\%, thus paving the way for advancements in automating quality assessment in clinical research.

\acknowledgements{%
We would like to thank the editorial committee of the TAL journal for inviting us to coordinate the scientific committee for this special issue.
%
We are especially grateful to the editors-in-chief for their support and invaluable assistance throughout this process.
Finally, we would like to thank all the reviewers and members of the scientific committee who joined us for this special issue and generously volunteered their time to help us select the articles published here. 
}

\bibliography{references}

\end{document}